\documentclass[11pt]{article}

\usepackage{geometry}
\usepackage{graphicx}
\graphicspath{{/home/ning/Dropbox/Research/Splines/plot/}}
\usepackage{amsmath, amssymb, amsfonts, amsthm, float,bm}  
\usepackage{enumerate,framed, multirow}
\usepackage[usenames, dvipsnames]{color}
\usepackage{comment, longtable, caption, subcaption, appendix}
\usepackage[sort&compress,longnamesfirst]{natbib}
\usepackage{setspace, parskip}
\usepackage{placeins}
\usepackage{url}

\binoppenalty=\maxdimen
\relpenalty=\maxdimen

\allowdisplaybreaks

\newcommand{\ppcite}[1]{\citeauthor{#1}' \citeyearpar{#1}}

\newtheorem{remark}{Remark}

\author{
Ning Dai\\ School of Statistics\\ University of Minnesota\\ \texttt{daixx224@umn.edu}
}
\title{Inference for Penalized Spline Regression:\\ Improving Confidence Intervals by Reducing the Penalty}
\date{\today}

\begin{document}

\maketitle

\begin{abstract}
Penalized spline regression is a popular method for scatterplot smoothing, but there has long been a debate on how to construct confidence intervals for penalized spline fits. Due to the penalty, the fitted smooth curve is a biased estimate of the target function. Many methods, including Bayesian intervals and the simple-shift bias-reduction, have been proposed to upgrade the coverage of the confidence intervals, but these methods usually fail to adequately improve the situation at predictor values where the function is sharply curved. In this paper, we develop a novel approach to improving the confidence intervals by using a smaller smoothing strength than that of the spline fits. With a carefully selected amount of reduction in smoothing strength, the confidence intervals achieve nearly nominal coverage without being excessively wide or wiggly. The coverage performance of the proposed method is investigated via simulation experiments in comparison with the bias-correction techniques proposed by \cite{Hodges} and \cite{KP15}.
\end{abstract}

\newpage

\section{Introduction}
A penalized spline is a non-parametric regression model commonly used for scatterplot smoothing, i.e., estimating a one-dimensional function of a single variable. Compared to other approaches to scatterplot smoothing, such as local polynomial fitting and classical series-based smoothers, penalized splines benefit from being a straightforward extension of linear regression modeling \cite[see][for examples and applications]{OS86, KellyRice90, Gray92, Gray94, EilersMarx96, Hastie96}. Asymptotic theories of penalized spline estimators have been explored over the last decade \citep{HO05,CKO09,KKF09,LiRuppert08,CW10,WSR11}.

The penalized spline smoother can be represented as a mixed linear model \citep{Robinson91, BrumbackRuppertWand99} and therefore aspects of mixed-linear-model theory can be applied to penalized splines. Most importantly, doing so enables the data to guide the choice of the amount of smoothing in a nearly automatic way \cite[Chapter 5.2]{RWC}, which has a profound influence on the fit. Apart from maximum likelihood-based smoothing parameter selection, there are more general methods based on classical model selection ideas that do not depend on the mixed model representation of penalized splines. The interested reader is referred to \citet[Chapter 5.3]{RWC} for a review of common approaches and \cite{AnsleyKohn85} and \cite{Wahba85} for examples. Compared to these methods, maximum likelihood-based smoothing parameter selection tends to be more robust and for a moderately misspecified correlation structure, over- or under-fitting does not occur \citep{KK07}.

Bayesian methods are also popular in scatterplot smoothing. The problem can be addressed using a Bayesian hierarchical model, in which a hyperprior for the smoothing parameter needs to be specified \cite[Chapter 16.3]{RWC}. Then the smoothing parameter, the spline coefficient, and an estimator of the target function are obtained as the posterior means. When a posterior mean is not available in closed form, Markov chain Monte Carlo is often used to approximately sample from the posterior distribution. On the other hand, an empirical Bayes approach enables a straightforward, data-driven way of choosing the smoothing parameter without requiring a hyperprior \citep{KP15}.

Let $f(x)$ denote the function that we want to estimate, i.e., the smooth curve representing the underlying trend of the scatterplot. With prespecified spline basis, degree, and knot locations, the smoothing parameter can be chosen by a variety of methods as introduced above. For a particular value $x$ of the predictor, the value of the scatterplot smooth, $\hat{f}(x)$, is a point estimate of $f(x)$. In this paper, we focus on inference for the unknown quantity $f(x)$ and more specifically, constructing a confidence interval for $f(x)$. We note that the intervals presented in this paper are all pointwise but not simultaneous. For a review of work on simultaneous confidence bands for penalized splines, we refer the interested reader to \citet[Chapter 6.5]{RWC} and \cite{KKC10}.

Constructing a confidence interval for $f(x)$ is a delicate matter. The main problem is the bias that is present in the point estimate $\hat{f}(x)$ due to the penalty. Various methods for constructing bias-corrected intervals have been proposed. Bayesian intervals \citep{Wahba83,Wood06,Weir97} incorporate the bias into the variance estimate. A frequentist interpretation of the adjustment was provided by \citet[pp.~139--140]{RWC} using the mixed model formulation of penalized splines. Bayesian intervals achieve good coverage averaged over the design points \citep{Nychka88} but provide no guarantee about pointwise coverage: As shown in \cite{RuppertCarroll00} and \cite{KP15}, if there are regions of sharp curvature in an otherwise flat regression function, then the coverage probability can be far below the nominal level in the regions of high curvature and greater than nominal elsewhere.

Instead of adjusting the variance, other methods focus on the inherently biased spline fit aiming to reduce the bias. \citet[pp.~96--98]{Hodges} proposed a simple-shift bias-correction method by subtracting an estimate of the bias from the fit. Improved coverage can be observed after the shift, but coverage is still poor at regions of high curvature due to residual bias. \cite{KP15} developed a similar but iterative technique that effectively reduces residual bias.

We aim to establish a method that performs well not only in flat regions but also at predictor values with sharp curvatures. Noticing that the unpenalized estimate is unbiased, we form a confidence interval based on the unpenalized fit. The unpenalized interval achieves good coverage at the cost of smoothness. To avoid an excessively wide or wiggly confidence band while maintaining the coverage level of the unpenalized method, a class of intervals is constructed with less severe penalization than that of the spline fit. With a carefully selected amount of reduction in smoothing strength, the proposed method outperforms \ppcite{Hodges} simple-shift correction, as we show via simulation experiments in Section \ref{sec:simulation}. In particular, we observe dramatic improvement at high curvatures and close-to-nominal coverage elsewhere. Compared to \ppcite{KP15} iterative procedure, the proposed method performs equally well, but is simpler to implement and more general.

The paper is structured as follows. We develop notation and background in Section \ref{sec:notation}. The proposed method is established in Section \ref{sec:unbiased}. We then briefly introduce current bias-correction methods in Section \ref{sec:current}. This is followed by simulation studies in Section \ref{sec:simulation} that examine the performance of the proposed confidence intervals in comparison with existing approaches. We close the paper with some concluding remarks in Section \ref{sec:conclude}.

\section{Notation and Background}
\label{sec:notation}
Consider smoothing a scatterplot where the data are denoted $(x_i,y_i)$ for $i=1,\ldots,n$. The underlying trend would be a function $f$ such as
\begin{equation}\label{eq:model}
y_i=f(x_i)+\epsilon_i,~\epsilon_i\overset{i.i.d.}{\sim}\mathrm{N}(0,\sigma ^2).
\end{equation}
The function $f$ is some unspecified ``smooth" function that needs to be estimated from the data. The function is modeled using a spline, that is,
\begin{equation}\label{eq:spline}
f(x)=\boldsymbol\beta^\top \boldsymbol B(x) ,
\end{equation}
where $\boldsymbol B(x)$ is the vector of known spline basis functions and $\boldsymbol\beta$ is the vector of unknown coefficients.

Let $\boldsymbol y=(y_1,y_2,\ldots,y_n)^\top$ and $\boldsymbol X$ be a matrix whose $i$th row is $\boldsymbol B(x_i)^\top$, the basis functions evaluated at $x_i$. Then a penalized spline fit is given by
\begin{equation}\label{eq:fit}
\hat{f}(x)=\hat{\boldsymbol\beta}^\top\boldsymbol B(x),
\end{equation}
where $\hat{\boldsymbol\beta}$ is the minimizer of
\begin{equation}\label{eq:cri}
\|\boldsymbol y-\boldsymbol X\boldsymbol\beta\|^2+\alpha \boldsymbol\beta^\top \boldsymbol D\boldsymbol\beta
\end{equation}
for some positive semidefinite matrix $\boldsymbol D$ and scalar $\alpha>0$. This has the solution
\begin{equation}\label{eq:solution}
\hat{\boldsymbol\beta}=(\boldsymbol X^\top\boldsymbol X+\alpha\boldsymbol D)^{-1}\boldsymbol X^\top\boldsymbol y.
\end{equation}

Taking $\alpha$ as known or given, the estimated smooth function $\hat{f}(x)$ evaluated at $x$ has the variance
\begin{equation}\label{eq:varfit}
\mathrm{Var}[\hat{f}(x)]=\sigma ^2 \boldsymbol B(x)^\top\left(\boldsymbol X^\top \boldsymbol X+\alpha\boldsymbol D\right)^{-1}\boldsymbol X^\top \boldsymbol X \left(\boldsymbol X^\top \boldsymbol X+\alpha\boldsymbol D\right)^{-1}\boldsymbol B(x).
\end{equation}
Then an approximate $100(1-\tau)\%$ CI is given by
\begin{equation}\label{eq:CI}
\hat{f}(x) \pm z_{\frac{\tau}{2}}\widehat{\mathrm{SD}}[\hat{f}(x)],
\end{equation}
where the standard error is obtained by replacing $\sigma ^2$ with some estimate $\hat{\sigma}^2$, which is uniquely determined by a preselected $\alpha$.

Suppose $\alpha^*$ is the optimal smoothing strength chosen by a particular criterion, say, maximum likelihood. Let $\hat{\boldsymbol\beta}_1$ and $\hat{f}_1$ denote, respectively, the estimated coefficients and smooth function when taking $\alpha=\alpha^*$. The interval \eqref{eq:CI} with $\alpha=\alpha^*$ is in common use despite the fact that its coverage probability often falls below the nominal level due to the inherent bias of the penalized estimate $\hat{f}_1(x)$.

\section{Improving Confidence Intervals by Reducing the Smoothing Strength}
\label{sec:unbiased}
As the unpenalized fit is unbiased, it can be used to construct confidence intervals with desirable coverage probabilities. Let
\begin{equation}\label{eq:Nmu}
\hat{\boldsymbol \beta}_0=(\boldsymbol X^\top \boldsymbol X)^{-1}\boldsymbol X^\top\boldsymbol y
\end{equation}
denote the unpenalized estimator of the coefficient $\boldsymbol \beta$, obtained from a spline regression without penalty, i.e., by letting $\alpha=0$ in the objective function \eqref{eq:cri}. Notice that $\hat{\boldsymbol \beta}_0$ is unbiased for $\boldsymbol \beta$ because $\mathrm{E}\boldsymbol y=\boldsymbol X\boldsymbol \beta$. Hence the unpenalized fit
\begin{equation}\label{eq:Nfit}
\hat{f}_0(x)=\hat{\boldsymbol \beta}_0^\top\boldsymbol B(x)
\end{equation}
is unbiased for $f(x)$.

The corresponding unpenalized interval outperforms a variety of alternatives with an impressive increase in coverage probabilities at predictor values with high curvatures, as shown in Section \ref{sec:simulation}. In fact, it will always have the best coverage because the unpenalized fit is the optimal solution of the bias-correction problem. However, the resulting pointwise confidence band is as wiggly as the unsmoothed fit and would thus, like the unsmoothed fit, be shunned by most users. Other than that, we know of no solid arguments against wiggly confidence bands.

There seems to be a trade-off between the coverage and the smoothness of the confidence band: The fully penalized spline fit yields a smooth confidence band with poor coverage; using the unpenalized estimate effectively improves the coverage at the cost of smoothness. To strike a balance between coverage and smoothness, we consider reducing the smoothing strength $\alpha^*$ by multiplying it by some scalar $\theta$ between 0 and 1, yielding the coefficient estimate
\begin{equation}\label{eq:Lmu}
\hat{\boldsymbol \beta}_\mathrm{\theta}=(\boldsymbol X^\top \boldsymbol X+\theta\alpha^*\boldsymbol D)^{-1}\boldsymbol X^\top\boldsymbol y
\end{equation}
and the corresponding estimator of $f(x)$
\begin{equation}\label{eq:Lfit}
\hat{f}_\mathrm{\theta}(x)=\hat{\boldsymbol \beta}_\mathrm{\theta}^\top\boldsymbol B(x).
\end{equation}
Notice that the original estimate and the unpenalized estimate are the two extreme cases where $\theta=1$ and $\theta=0$, respectively.

Another advantage of $\hat{f}_\mathrm{\theta}(x)$ with $0<\theta<1$ compared to the unpenalized fit is that it generates narrower intervals than the unpenalized fit. Recall that the width of a confidence interval depends on the amount of smoothing. As can be observed in \eqref{eq:varfit}, increased smoothing strength leads to reduced variance of the point estimator and hence a narrower confidence interval.

To conclude, when coverage is the major concern, one should use the unpenalized confidence interval. If wide or wiggly intervals are to be avoided, we suggest constructing a confidence interval based on the less penalized fit $\hat{f}_\mathrm{\theta}(x)$ with a carefully selected value of $\theta$.

\section{Current Bias-Correction Methods}
\label{sec:current}
In this section we briefly review two bias-correction methods proposed by \citet[pp.~96--98]{Hodges} and \cite{KP15}. Both approaches effectively reduce the bias in the fully penalized spline fit $\hat{f}_1(x)$, leading to confidence intervals with improved coverage.

Recall that the fully penalized solution $\hat{\boldsymbol\beta}_1$ has the inherent bias \cite[p.~139]{RWC}
\begin{equation}\label{eq:bias}
\mathrm{bias}[\hat{\boldsymbol\beta}_1]=\mathrm{E}[\hat{\boldsymbol\beta}_1]-\boldsymbol\beta=-\alpha^*\left(\boldsymbol X^\top \boldsymbol X+\alpha^*\boldsymbol D\right)^{-1}\boldsymbol D\boldsymbol\beta,
\end{equation}
which is unknown and needs to be estimated from the data. The quality of the bias estimation is crucial. If an appropriate estimate of the bias is available, then a less biased estimator of $\boldsymbol\beta$ can be obtained by subtracting the estimated bias from $\hat{\boldsymbol\beta}_1$. On the other hand, if the bias is estimated poorly, then subtracting it just adds noise without improving the coverage \citep{SunLoader}.

\citet[pp.~96--98]{Hodges} proposed a simple-shift correction method where the bias is estimated by substituting $\boldsymbol\beta$ in \eqref{eq:bias} with $\hat{\boldsymbol\beta}_1$. \cite{Hodges} then formed a confidence interval with the new fit and the variance estimate of the original fit $\hat{f}_1(x)$. \cite{SunLoader} and \cite{Cunanan} showed that this interval displays lower coverage probability than both the fully penalized interval and the Bayesian interval at linear points of a curve and performs only slightly better when the bias is larger than the variance. Fortunately, this problem can be easily fixed by using the variance estimate of the corrected fit instead of that of the original fit. Henceforth, ``\ppcite{Hodges} method" refers to constructing confidence intervals with the updated variance estimates.

\cite{KP15} provided a procedure that iteratively updates the estimated bias. If only one iteration is done, \ppcite{KP15} approach is equivalent to \ppcite{Hodges}. We now formally present \ppcite{KP15} method.

\cite{KP15} established an iteratively bias-corrected bootstrap technique for constructing improved confidence intervals in an attempt to solve the high energy physics unfolding problem in which the goal is to estimate the spectrum of elementary particles given observations distorted by the limited resolution of a particular detector, or, more precisely, to estimate the intensity function of an indirectly observed Poisson point process.

This iterative bootstrap procedure was originally developed for the unfolding problem. When adapted to the penalized spline model with normal errors \eqref{eq:model}, the bias is available in closed form, and thus bootstrapping is unnecessary for bias estimation. We now describe their method in the context of penalized spline regression with normal errors and derive its asymptotic behavior w.r.t. the number of iterations, i.e., the iteratively updated interval will eventually converge to the unpenalized interval. 

Let $N_\mathrm{BC}$ be the number of bias-correction iterations. Starting with the fully penalized estimate $\hat{\boldsymbol\beta}^{(0)}=\hat{\boldsymbol\beta}_1$, for $i=0$ to $N_\mathrm{BC}-1$ do
\begin{enumerate}
\item Estimate the bias by replacing $\boldsymbol\beta$ with $\hat{\boldsymbol\beta}^{(i)}$ in \eqref{eq:bias}: $\widehat{\mathrm{bias}}^{(i)}[\hat{\boldsymbol\beta}_1]
=-\alpha^*\left(\boldsymbol X^\top \boldsymbol X+\alpha^*\boldsymbol D\right)^{-1}\boldsymbol D\hat{\boldsymbol\beta}^{(i)}$;\label{step3}
\item Set $\hat{\boldsymbol\beta}^{(i+1)}=\hat{\boldsymbol\beta}_1-\widehat{\mathrm{bias}}^{(i)}[\hat{\boldsymbol\beta}_1]$.\label{step4}
\end{enumerate}
Return $\hat{\boldsymbol\beta}_\mathrm{BC}=\hat{\boldsymbol\beta}^{(N_\mathrm{BC})}$.

The bias-corrected spline coefficients $\hat{\boldsymbol\beta}_\mathrm{BC}$ are associated with a bias-corrected function estimate $\hat{f}_\mathrm{BC}(x)=\hat{\boldsymbol\beta}_\mathrm{BC}^\top \boldsymbol B(x)$. Then the variability of the bias-corrected estimator $\hat{f}_\mathrm{BC}(x)$ is used to construct confidence intervals for $f(x)$. Specifically, an approximate $100(1-\tau)\%$ CI is given by
\begin{equation}\label{eq:CI_BC}
\hat{f}_\mathrm{BC}(x) \pm z_{\frac{\tau}{2}}\widehat{\mathrm{SD}}[\hat{f}_\mathrm{BC}(x)]\, .
\end{equation}

By induction we obtain for all $i$,
\[\hat{\boldsymbol\beta}^{(i)}=\sum_{j=0}^i M^j\hat{\boldsymbol\beta}_1,\]
where $M=\alpha^*\left(\boldsymbol X^\top \boldsymbol X+\alpha^*\boldsymbol D\right)^{-1}\boldsymbol D$.
In particular,
\[\hat{\boldsymbol\beta}_\mathrm{BC}=\sum_{j=0}^{N_\mathrm{BC}} M^j\hat{\boldsymbol\beta}_1.\]
By letting the number of iterations $N_\mathrm{BC}\rightarrow\infty$, we obtain $\hat{\boldsymbol\beta}_\mathrm{BC}\rightarrow(I-M)^{-1}\hat{\boldsymbol\beta}_1$.
That is, as $N_\mathrm{BC}\rightarrow\infty$,
\[
\hat{\boldsymbol\beta}_\mathrm{BC}\rightarrow\hat{\boldsymbol\beta}_0.
\]

We have demonstrated that the result of the iterative bias correction converges to the unpenalized spline fit as the number of iterations tends to infinity. \cite{KP15} showed via a simulation study that ``a single bias-correction iteration already improves the coverage significantly, with further iterations always improving the performance". This suggests that the unpenalized interval as the limit of the iterative approximation has better coverage than \ppcite{KP15} interval of any number of iterations, including \ppcite{Hodges} shifted interval which is the case when $N_\mathrm{BC}=1$. However, instead of requiring a large $N_\mathrm{BC}$, \cite{KP15} ``preferred settling with $N_\mathrm{BC}=5$, as increasing the number of iterations produced increasingly wiggly intervals".
\section{Simulation Experiments}
\label{sec:simulation}
\subsection{An Example}
\label{sub:real}
We implement the proposed method on the \verb+fossil+ data, available in the R package \verb+SemiPar+. The data frame has 106 observations on fossil shells taken on two variables -- the predictor, \verb+age+, in millions of years, and the response, \verb+strontium.ratio+, ratios of strontium isotopes.

We fit a penalized regression spline having 26 equally spaced knots within the range of \verb+age+ as recommended by \citet[p.~126]{RWC}. An order-4 B-spline basis is used with the smoothness penalty as the integrated square of the second derivative \citep{OS86,OS88}, i.e.,
\begin{equation}\label{eq:penalty}
\boldsymbol D=\int \boldsymbol B''(x)\boldsymbol B''(x)^\top dx.
\end{equation}
The smoothing parameter $\alpha=\alpha^*$ is selected by REML using the mixed model representation. We then construct nominal 95\% confidence intervals using the proposed method of various $\theta$ values and plot the confidence bands.

\begin{figure}[t!]
    \centering
        \begin{subfigure}[t]{.5\textwidth}
        \centering
        \includegraphics[width=\linewidth]{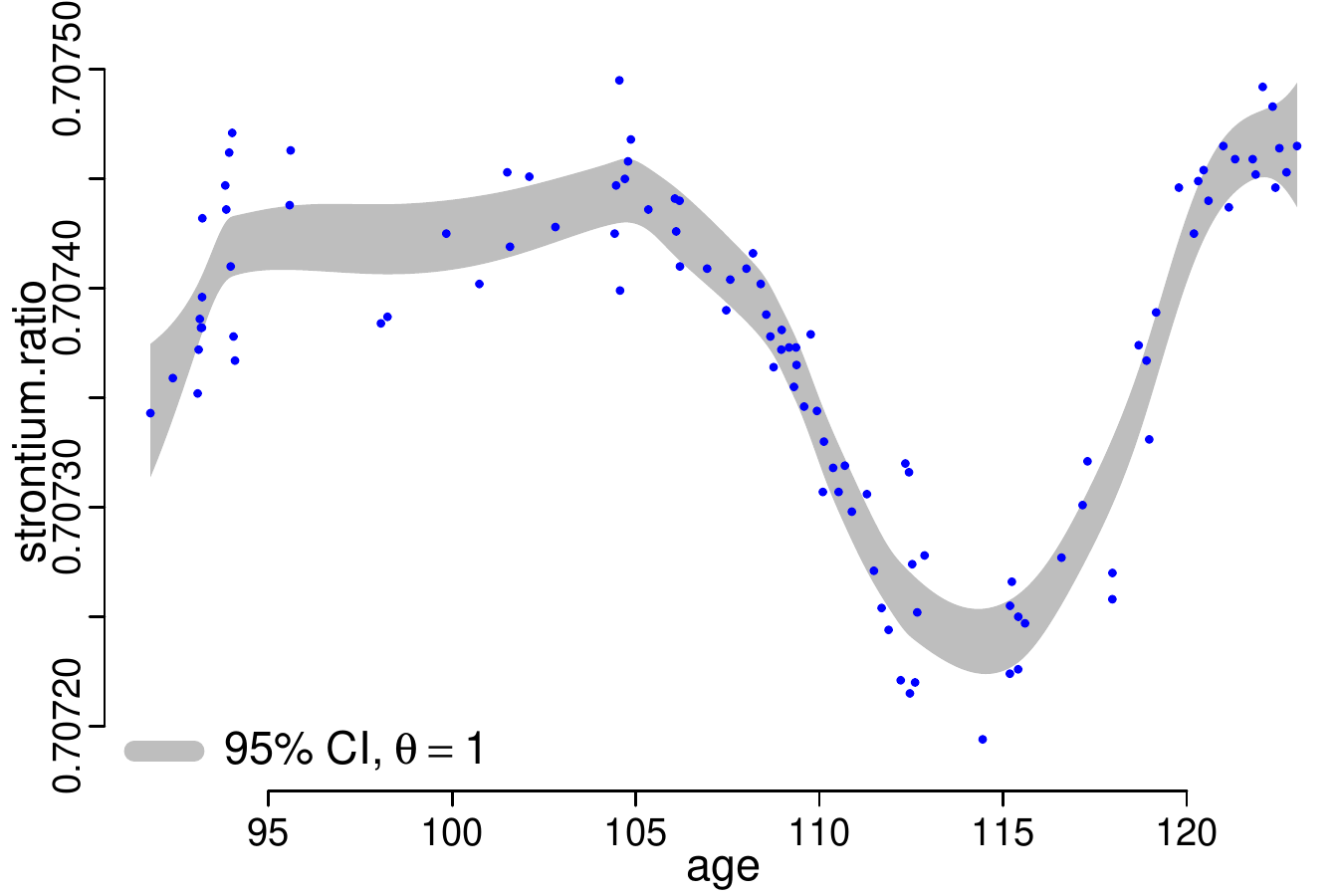}
        \caption{$\theta=1$. Fully penalized.}\label{fig:fos1}
    \end{subfigure}
    ~
    \begin{subfigure}[t]{.5\textwidth}
        \centering
        \includegraphics[width=\linewidth]{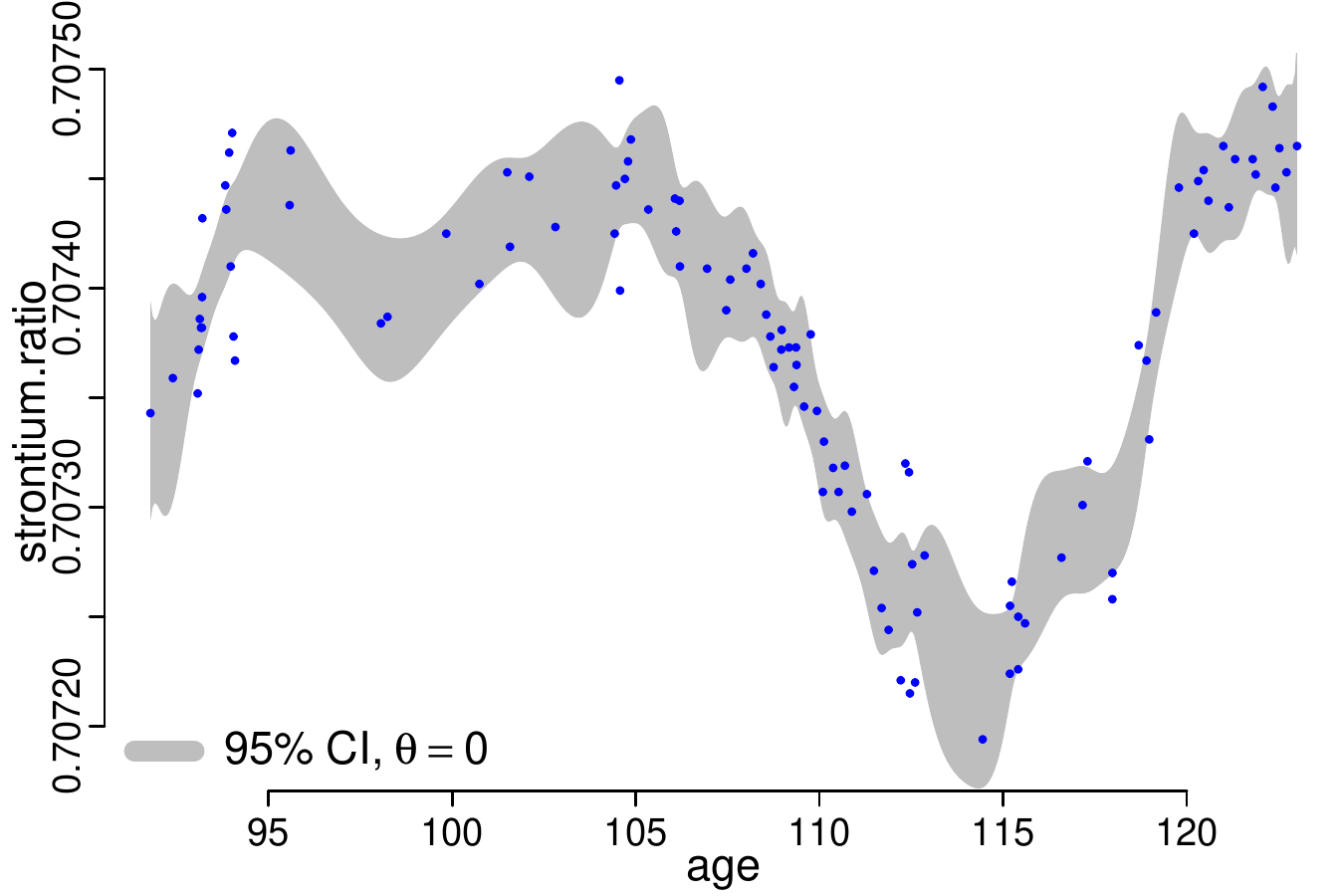}
        \caption{$\theta=0$. Unpenalized.}\label{fig:fos0}
    \end{subfigure}%
    ~ 
    \begin{subfigure}[t]{.5\textwidth}
        \centering
        \includegraphics[width=\linewidth]{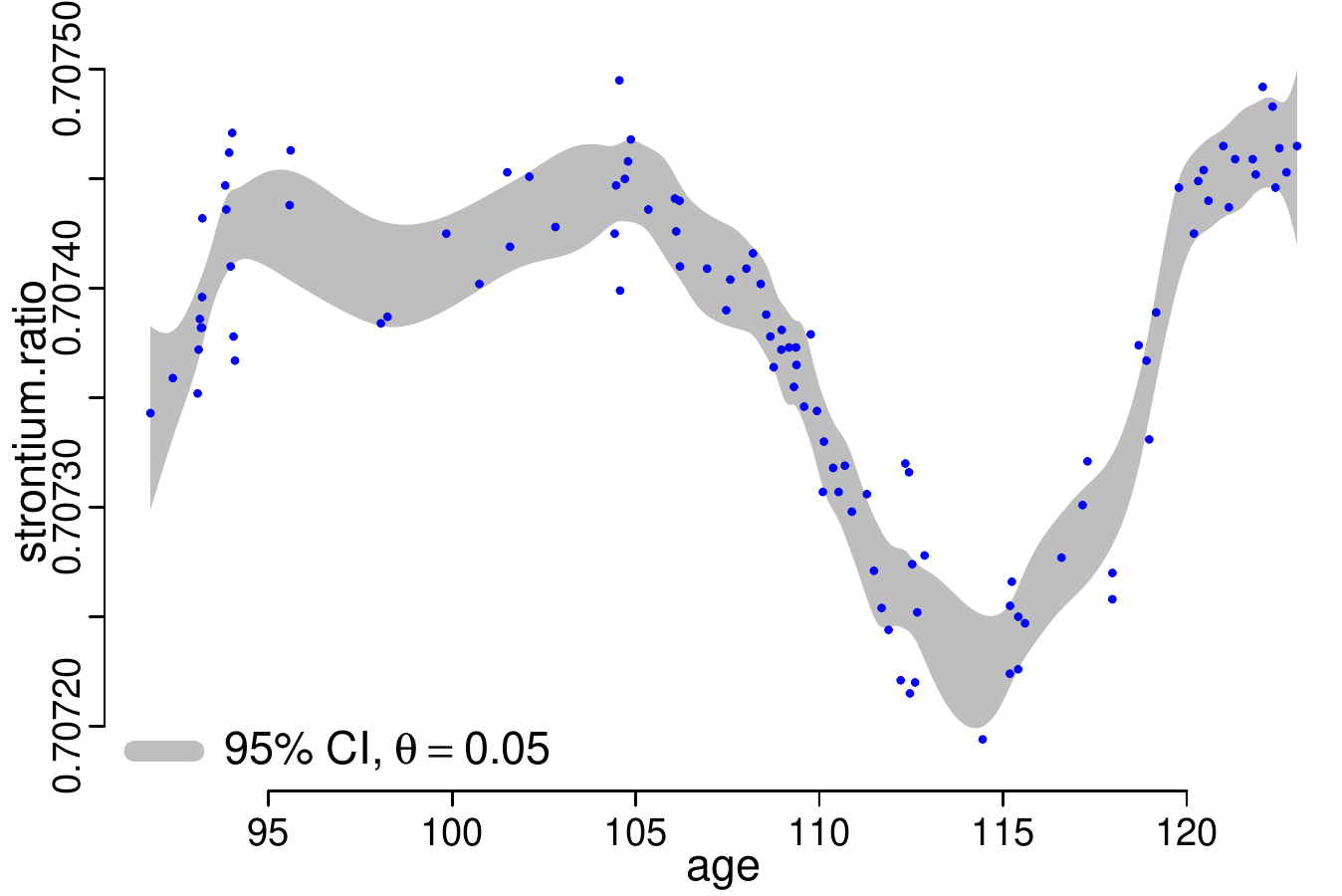}
        \caption{$\theta=0.05$.}\label{fig:fos005}
    \end{subfigure}
    ~
    \begin{subfigure}[t]{.5\textwidth}
        \centering
        \includegraphics[width=\linewidth]{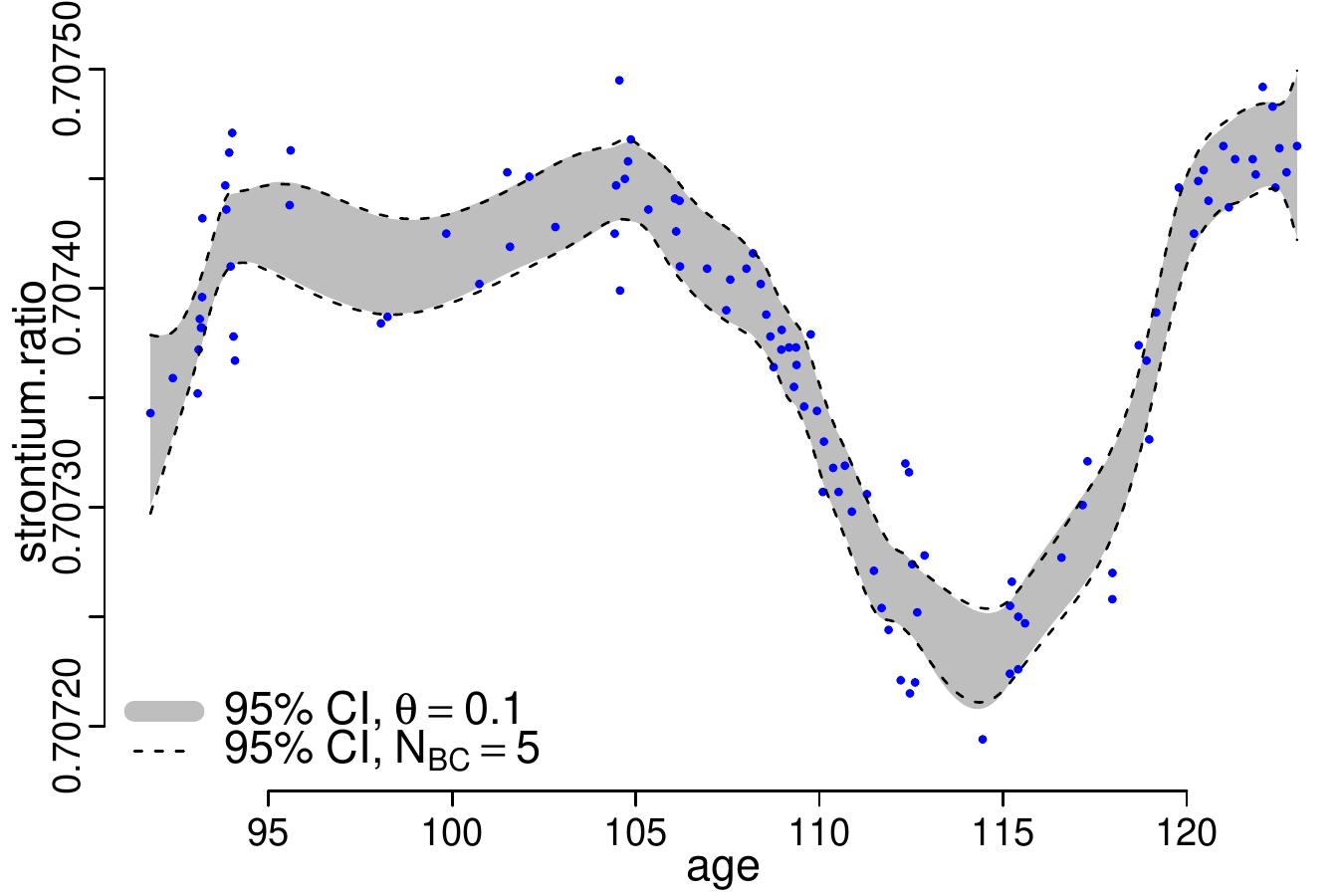}
        \caption{$\theta=0.1$. With imposed iteratively corrected CI using 5 iterations.}\label{fig:fos01}
    \end{subfigure}%
    ~
    \begin{subfigure}[t]{.5\textwidth}
        \centering
        \includegraphics[width=\linewidth]{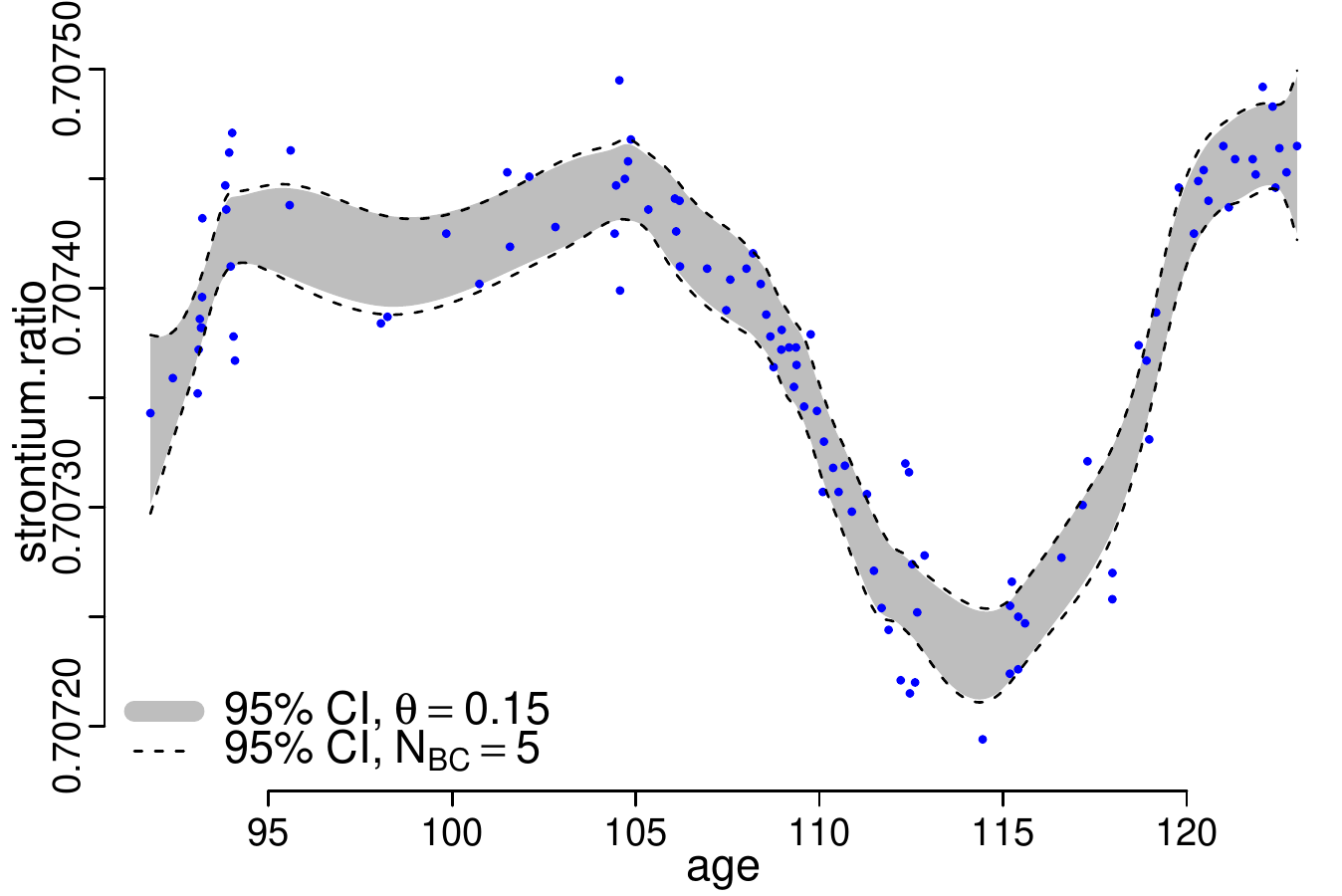}
        \caption{$\theta=0.15$. With imposed iteratively corrected CI using 5 iterations.}\label{fig:fos015}
    \end{subfigure}
    \caption{Nominal $95\%$ confidence bands produced by the proposed method (gray regions) using different $\theta$ values ($\theta\in\{0,0.05,0.1,0.15\}$) are compared with the fully penalized confidence band ($\theta=1$). When $\theta=0.1$ and $0.15$, the resulting confidence band is close to that of the iterative bias-correction method using 5 iterations (dashed lines).}
    \label{fig:fos}
\end{figure}

Recall that $\theta$ is the ratio of the smoothing strength of the confidence band to that of the smoothed curve. We considered different values of $\theta$: $\theta=0$, corresponding to the unpenalized interval; $\theta=0.05,0.1,0.15$, corresponding to less penalized interval; $\theta=1$, which is the fully penalized interval. The resulting confidence bands are compared in Figure \ref{fig:fos}.

The fully penalized confidence band with $\theta=1$ (Figure \ref{fig:fos1}) is narrow and smooth. When a smaller penalty is used to form the confidence intervals, we obtain a wigglier and wider confidence band. The unpenalized confidence band with $\theta=0$ (Figure \ref{fig:fos0}) appears extremely wiggly and the interval at each predictor value is much wider than those for $\theta\in\{0.05,0.1,0.15\}$. When $\theta$ increases slightly from 0 to $0.05$, we observe a substantially smoother and narrower confidence band (Figure \ref{fig:fos005}). Figures \ref{fig:fos01} and \ref{fig:fos015} show that further increases in $\theta$ yield still smoother and narrower confidence intervals. We also notice that the proposed method using $\theta=0.1$ and $0.15$ generates confidence bands very similar to those of the iterative bias-correction method with 5 iterations.

\subsection{Investigating Coverage Performance}
\label{sub:coverage}
Our goal is to investigate through simulation experiments the coverage performance of the proposed confidence intervals constructed with reduced smoothing strength. Here we show selected results for one simulation setting; complete results are in the supplement \cite[Section 1]{Dai}.

We consider examples with varying degrees of curvature, including very high curvature where existing methods for constructing confidence intervals have major problems with undercoverage, precisely so we can compare the new and old methods in the places where the old methods perform least well.

The true data generation function, shown in Figures \ref{fig:CI005} and \ref{fig:CIband}, is a variant of the ``broken-stick" function having sudden turns at $x=1$ and $x=3$ \citep{Cunanan}. Since sharp corners are rarely seen in practice, we smooth out the edges with quadratic curves tangent to the ``broken-stick" function at $x=0.8,1.2$ and $x=2.8,3.2$ to imitate artifacts such as fast turns or long valleys seen in real data. The exact function is stated and plotted in the supplement \cite[Section 1.1]{Dai}.

We first demonstrate the proposed method using simulated data. A dataset of $n=101$ observations was generated using a single predictor. We carried out the simulation through the following procedure.
\begin{enumerate}[(a)]
\item Take $n=101$ equally spaced values $\{x_1,x_2,\ldots,x_n\}$ on $[0,5]$.
\item Generate $n$ i.i.d. error terms $\{\epsilon_1,\epsilon_2,\ldots,\epsilon_n\}$ from $N(0,\sigma^2)$, where $\sigma=0.1$.
\item Compute each $y_i$ by adding the error term $\epsilon_i$ to $f(x_i)$.
\item Fit a penalized regression spline having 24 equally spaced knots between 0 and 5 as recommended by \citet[p.~126]{RWC}. An order-4 B-spline basis is used with the smoothness penalty \eqref{eq:penalty}.
The smoothing parameter $\alpha=\alpha^*$ is selected by REML using the mixed model representation.
\item Construct a nominal 95\% confidence band.
\end{enumerate}

\begin{figure}[t!]
    \centering
    \begin{subfigure}[t]{.75\textwidth}
        \centering
        \includegraphics[width=\linewidth]{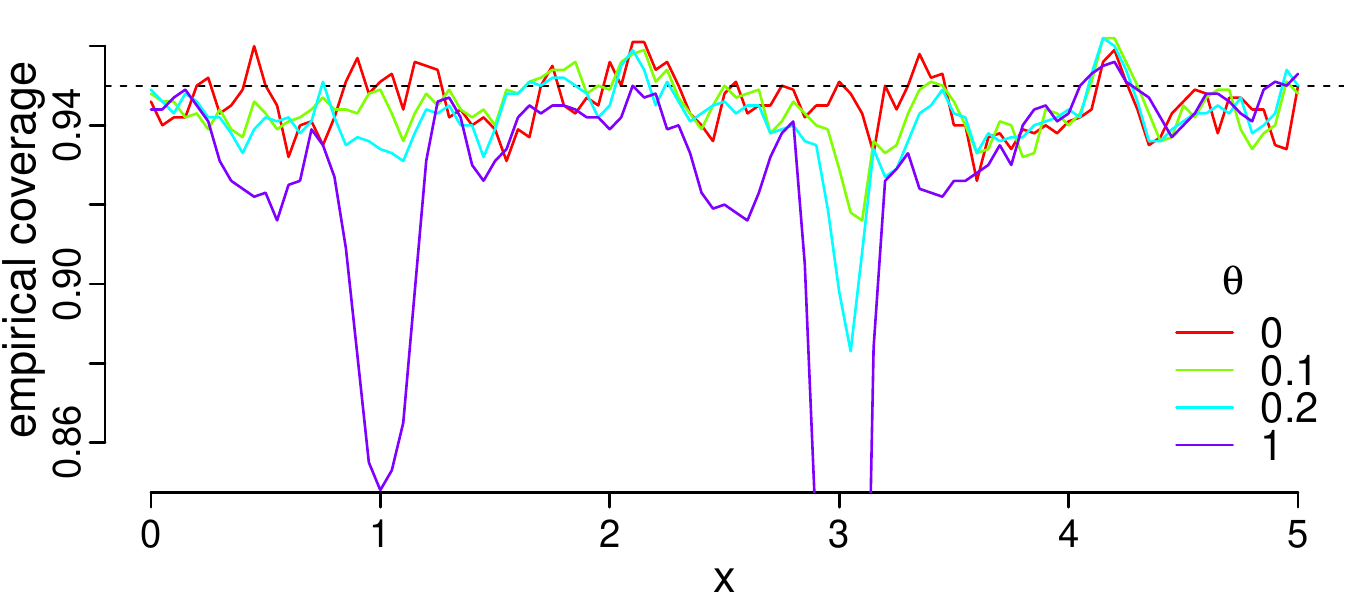}
        \caption{Reducing the smoothing strength improves the coverage. The fully penalized confidence intervals ($\theta=1$) have the worst coverage, while the unpenalized and thus effectively unbiased confidence intervals ($\theta=0$) perform the best.}\label{fig:RBthe}
    \end{subfigure}
    ~
\begin{subfigure}[t]{.75\textwidth}
        \centering
        \includegraphics[width=\linewidth]{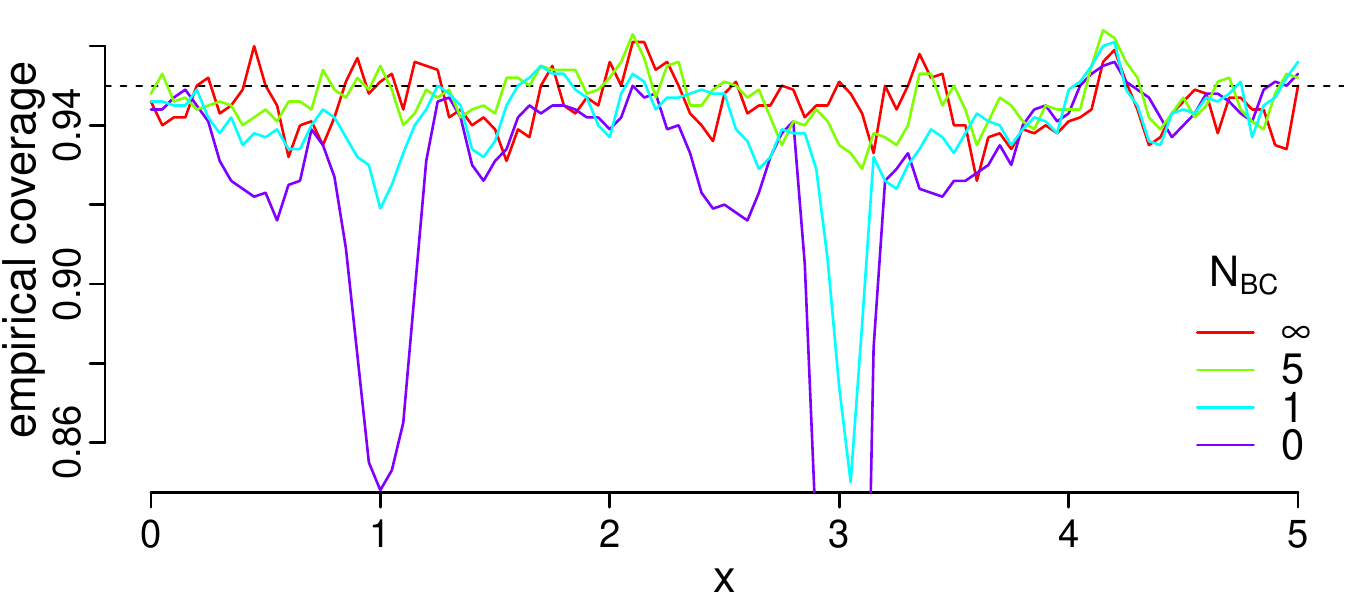}
        \caption{Bias-correction iterations improve the coverage. The fully penalized confidence intervals ($N_\mathrm{BC}=0$) have the worst coverage, while the unpenalized and thus effectively unbiased confidence intervals ($N_\mathrm{BC}=\infty$) perform the best.}\label{fig:RBnit}
    \end{subfigure}
    \caption{Empirical coverage probabilities of nominal $95\%$ confidence intervals produced by the proposed method and the iterative bias-correction method. Lowest coverage of the fully penalized confidence band ($\theta=1$ and $N_\mathrm{BC}=0$) is $0.654$ at $x=3.05$.}
    \label{fig:RB}
\end{figure}

\begin{figure}[t!]
    \centering
        \begin{subfigure}[t]{.75\textwidth}
        \centering
        \includegraphics[width=\linewidth]{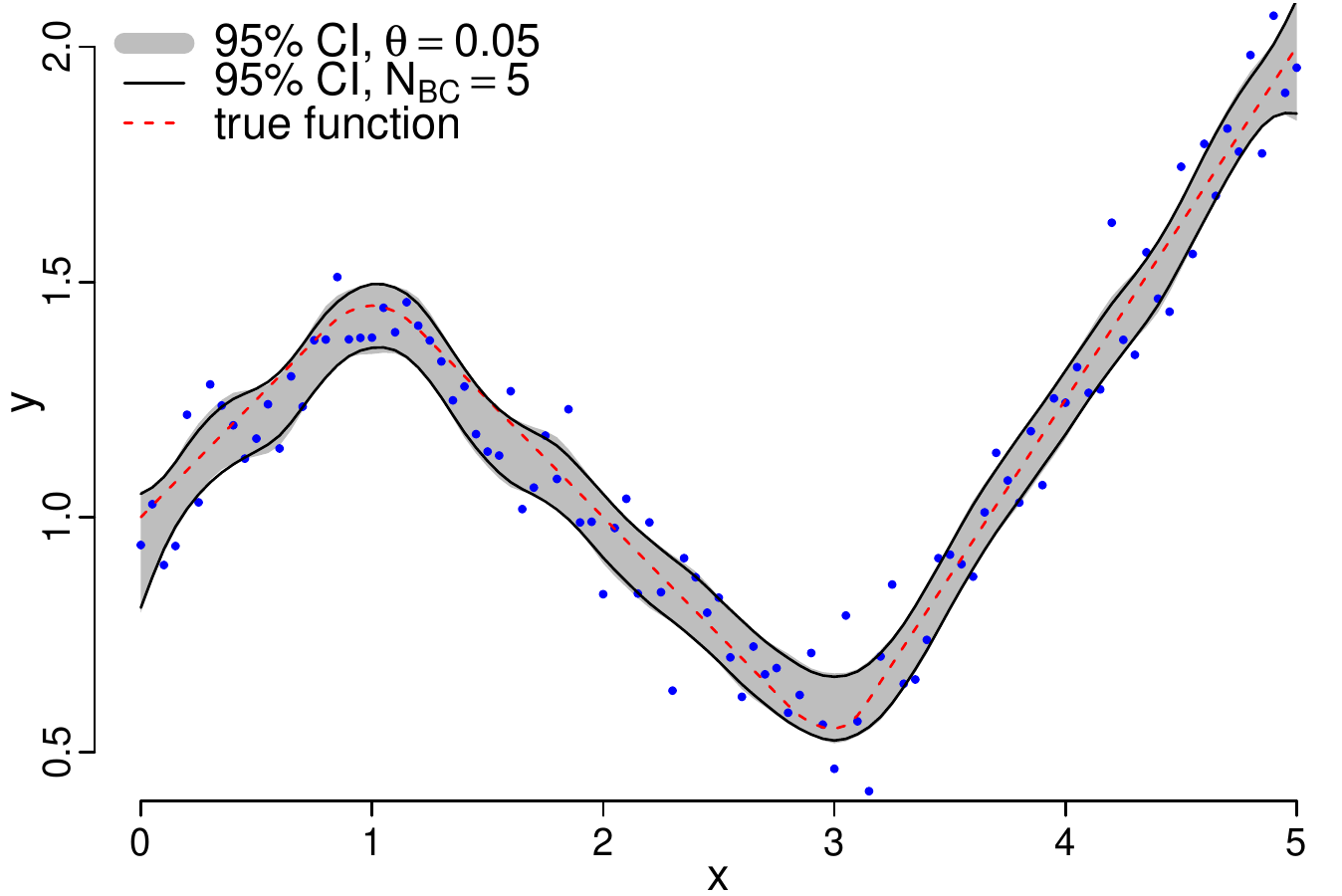}
        \caption{Nominal $95\%$ confidence bands produced by the proposed method with $\theta=0.05$ and the iterative bias correction with $N_\mathrm{BC}=5$.}\label{fig:CI005}
    \end{subfigure}
    ~
    \begin{subfigure}[t]{.75\textwidth}
        \centering
        \includegraphics[width=\linewidth]{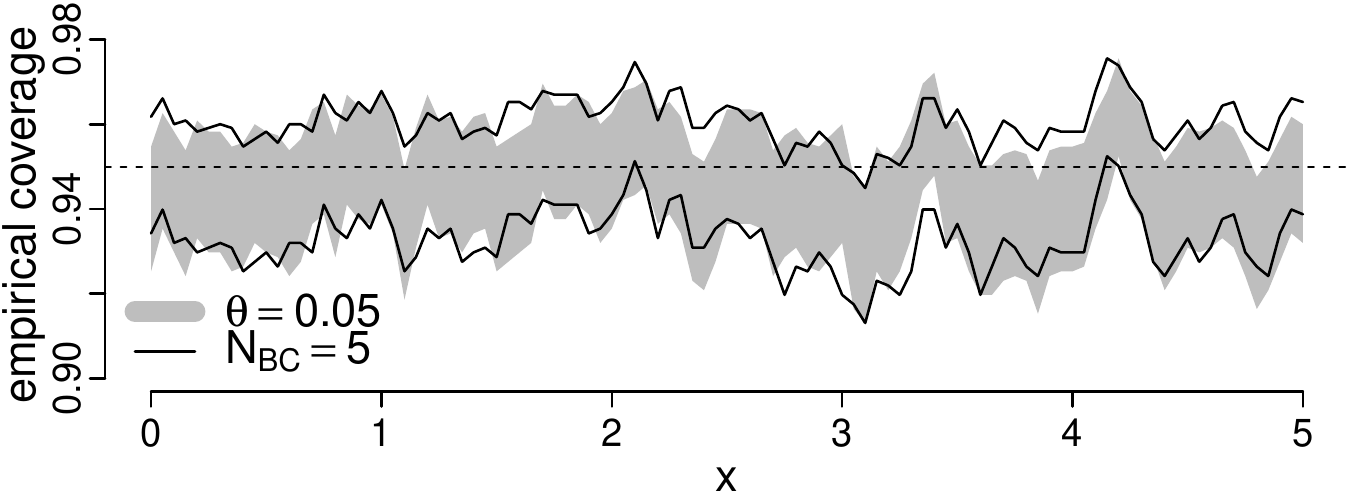}
        \caption{Empirical coverage probabilities (plus and minus $1.96$ times the Monte Carlo errors) of nominal $95\%$ confidence intervals produced by the proposed method with $\theta=0.05$ and the iterative bias correction with $N_\mathrm{BC}=5$.}\label{fig:comp}
    \end{subfigure}
    \caption{The proposed method with $\theta=0.05$ and the iterative bias correction with $N_\mathrm{BC}=5$ yield comparable confidence bands and coverage performance.}
    \label{fig:thenit}
\end{figure}

\begin{figure}[t!]
    \centering
        \begin{subfigure}[t]{.5\textwidth}
        \centering
        \includegraphics[width=\linewidth]{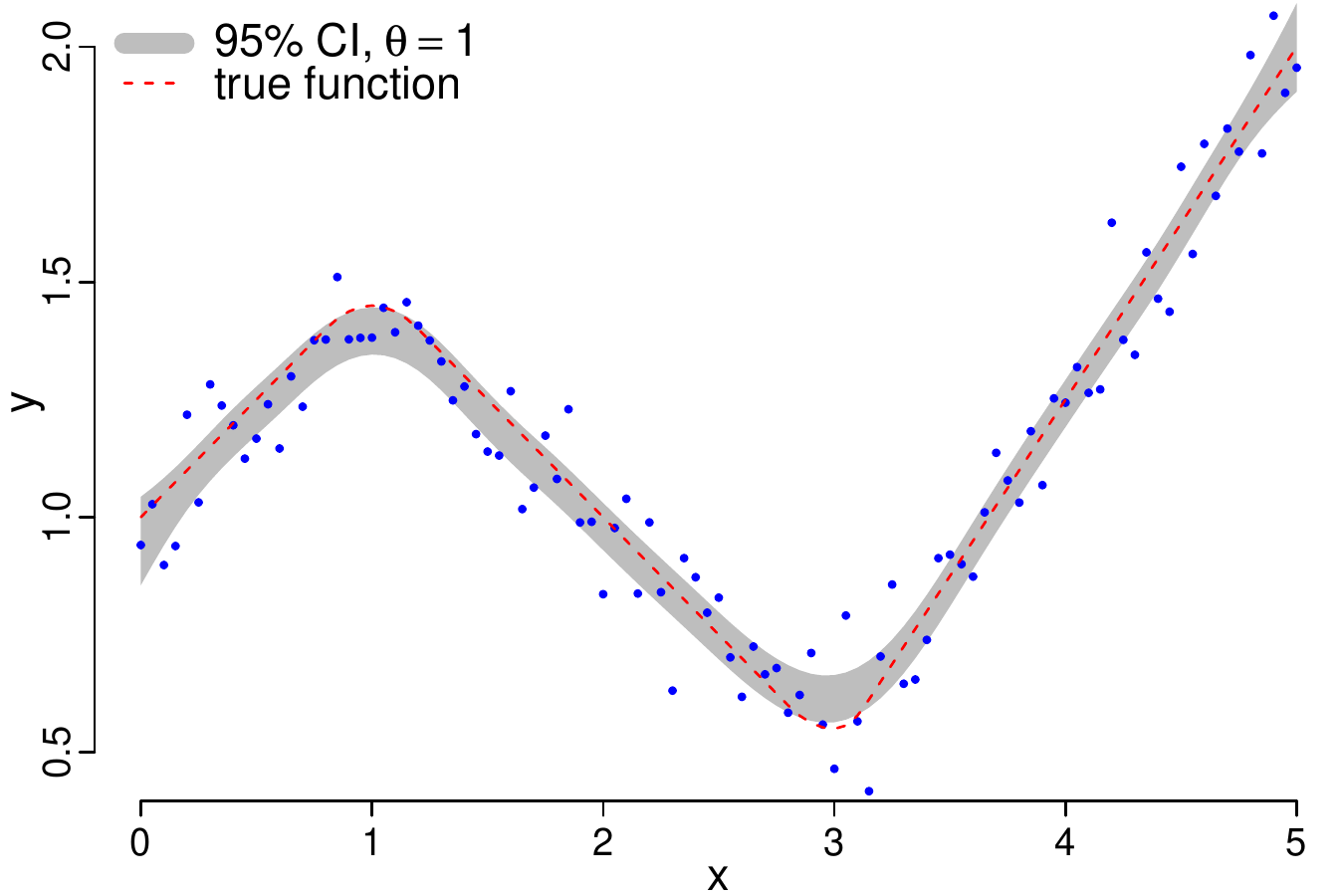}
        \caption{$\theta=1$. Fully penalized.}\label{fig:CI1}
    \end{subfigure}%
    ~
    \begin{subfigure}[t]{.5\textwidth}
        \centering
        \includegraphics[width=\linewidth]{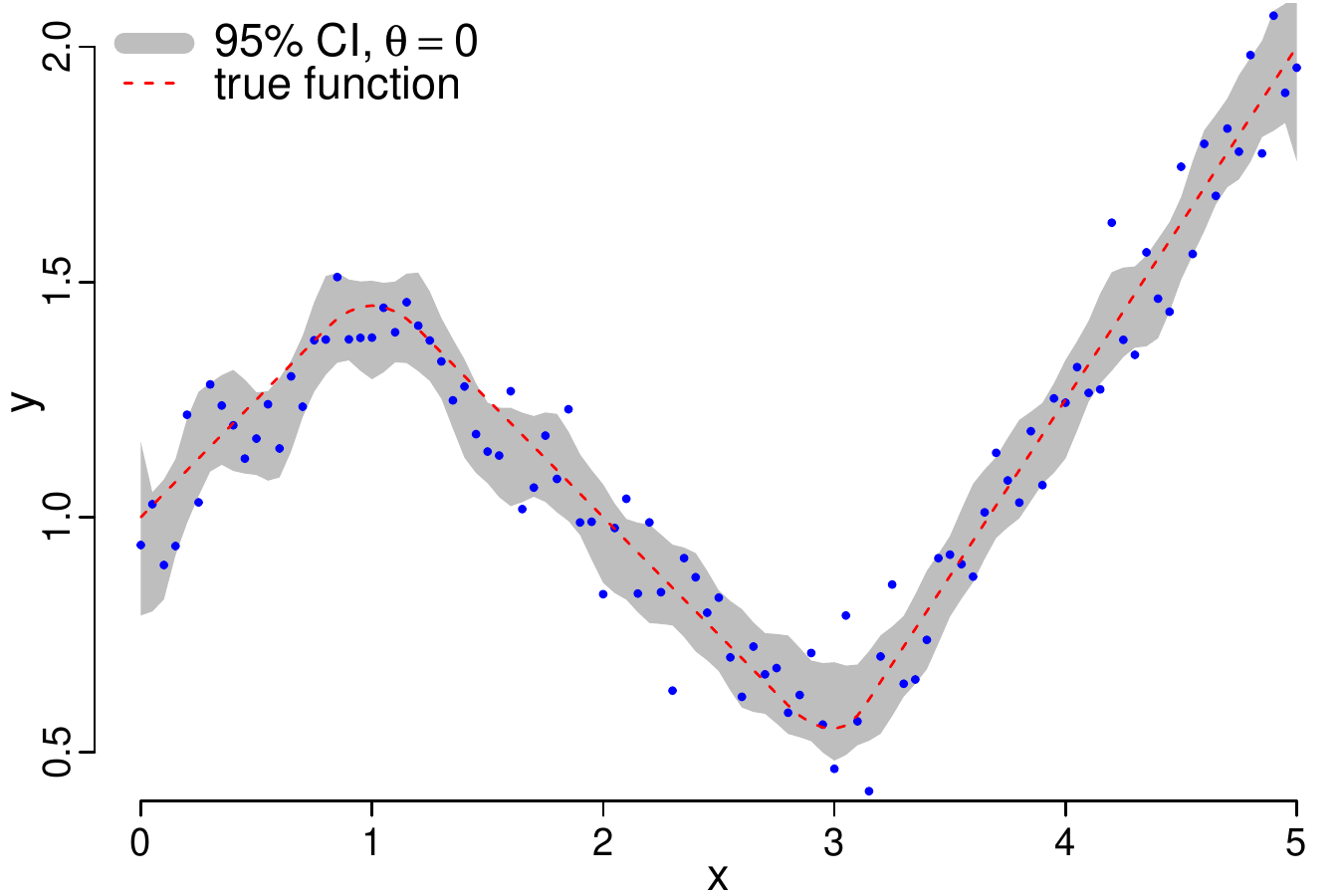}
        \caption{$\theta=0$. Unpenalized.}\label{fig:CI0}
    \end{subfigure}
    ~ 
    \begin{subfigure}[t]{.75\textwidth}
        \centering
        \includegraphics[width=\linewidth]{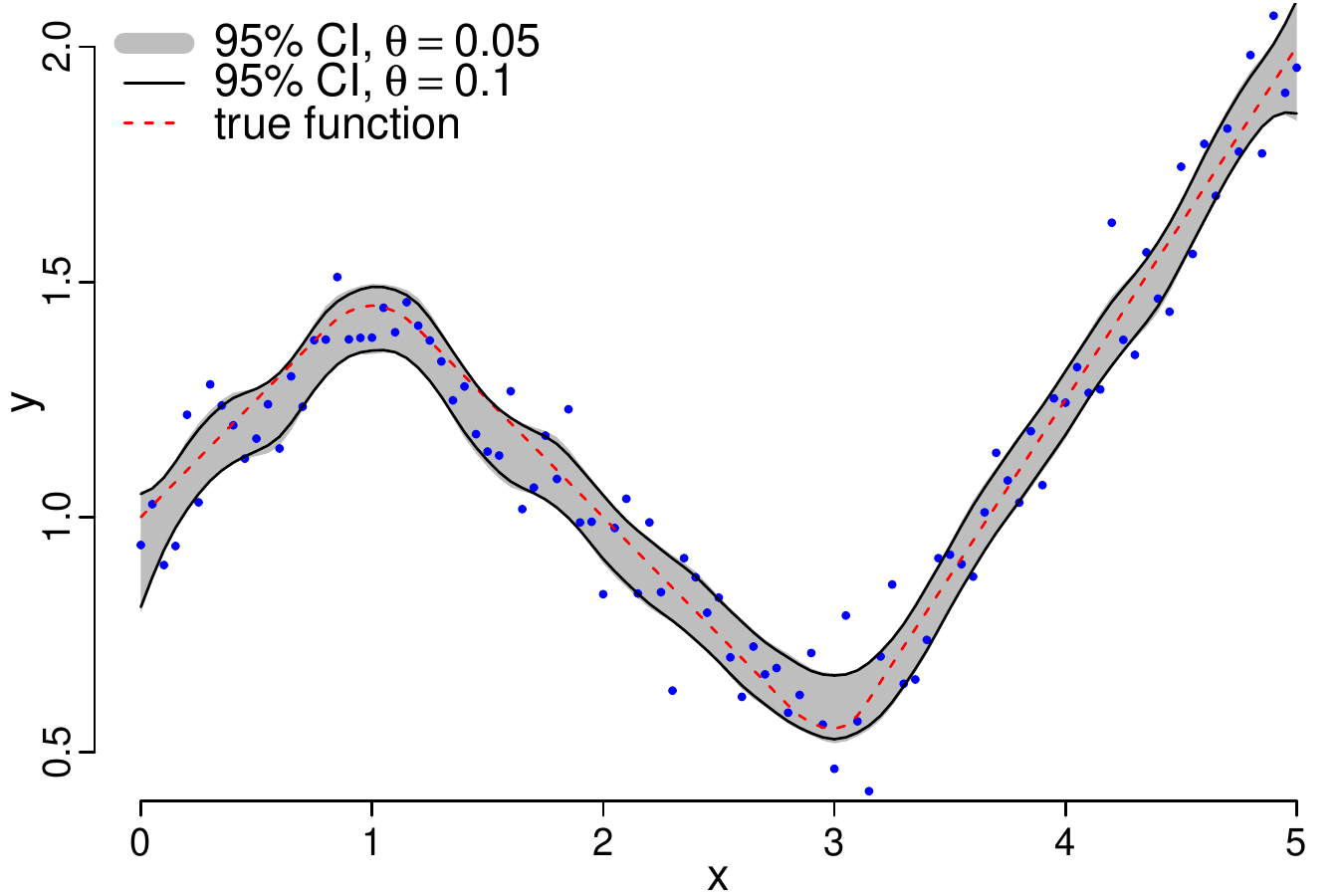}
        \caption{$\theta=0.05$ and $\theta=0.1$.}\label{fig:CIcomp}
    \end{subfigure}
    \caption{Nominal $95\%$ confidence bands using $\theta=0,0.05,0.1,1$.}
    \label{fig:CIband}
\end{figure}

We performed an empirical coverage study for the proposed method using $\theta=0,0.1,0.2$ in comparison with the iterative bias-correction approach developed by \cite{KP15}. We implemented \ppcite{KP15} method with $N_\mathrm{BC}=1$, which is equivalent to the simple-shift correction by \cite{Hodges}, and $N_\mathrm{BC}=5$, as recommended by \cite{KP15}. To assess the coverage probabilities, we repeated the simulation steps (a)-(e) 1000 times and calculated the rate at which the confidence intervals covered the true function at each predictor value. The results are reported in Figure \ref{fig:RB}.

Figure \ref{fig:RBthe} shows the empirical coverage of the proposed method that forms confidence intervals with reduced smoothing strength. The fully penalized intervals with $\theta=1$ are computed with the smoothing parameter value selected by the REML criterion. This data-guided smoothing strength produces a smooth curve estimate with inherent bias, leading to a confidence band that usually undercovers at regions having sizable bias, such as near $x=1$ and $x=3$. When the smoothing strength decreases by $80\%$, i.e., $\theta=0.2$, coverage probabilities have already increased significantly, with further reduction in the smoothing strength always improving the performance. The unpenalized confidence band with $\theta=0$ displays close-to-nominal coverage for all values of $x$.

Figure \ref{fig:RBnit} demonstrates the effect of the iterative bias correction established by \cite{KP15}. We observe improved coverage rates as the number of iterations increases. The iteratively updated confidence band with $N_\mathrm{BC}=1$ achieves comparable coverage to the proposed method using $\theta=0.2$ in flat regions and lower coverage in a small section near $x=3$ and another one near $x=1$. When $N_\mathrm{BC}=5$, it is comparable to the proposed method with $\theta$ between 0 and $0.1$. Figure \ref{fig:thenit} shows that $N_\mathrm{BC}=5$ and $\theta=0.05$ result in similar confidence bands as well as coverage probabilities.

The proposed method and the iterative bias-correction approach are essentially similar as both use less biased curve estimates to construct confidence intervals. With carefully selected $\theta$ and $N_\mathrm{BC}$, both methods generate confidence intervals that perform well at covering the target function without being extremely wiggly or wide.

An advantage of the proposed method over \ppcite{KP15} is that it is easier to pick an appropriate value of $\theta$ than $N_\mathrm{BC}$. We note that \cite{KP15} recommended $N_\mathrm{BC}=5$ based only on the results of their experiments. Due to a lack of knowledge of the theoretical properties of an iteration of their procedure, it is impossible to provide general guidance on how to choose $N_\mathrm{BC}$ in such a way that it optimizes the smoothness and width of the intervals while maintaining the coverage. The smoothness of the iterative result depends not only on $N_\mathrm{BC}$ but also on the rate of convergence of the iterative process. In contrast, when using the proposed method, one has full control over the degree of smoothness by selecting $\theta$. Therefore, to pick an appropriate value of $\theta$, one can preselect a desired range of smoothing strength and compare the resulting coverage probabilities using $\theta$ values such that $\theta\alpha^*$ varies within the range.

\begin{figure}[t!]
    \centering
    \includegraphics[width=.75\linewidth]{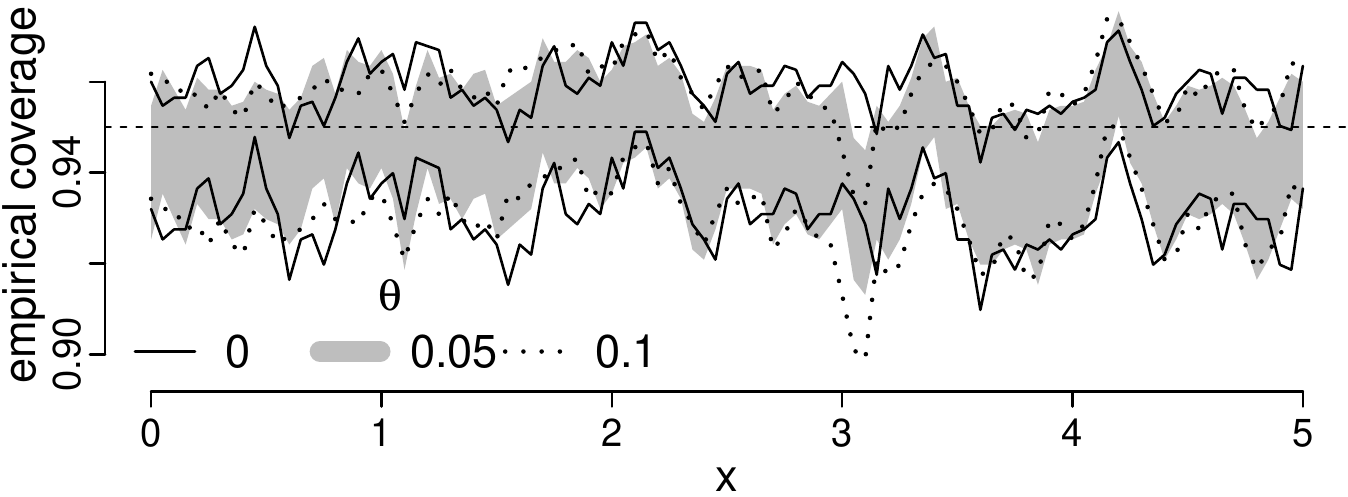}
    \caption{Investigating coverage performance: Empirical coverage probabilities (plus and minus $1.96$ times the Monte Carlo errors) of nominal $95\%$ confidence intervals produced by the proposed method using $\theta=0,0.05,0.1$. As $\theta$ increases from 0 to $0.1$, coverage level is generally preserved across the range of $x$, except for a small region near $x=3$.}
    \label{fig:CPthe}
\end{figure}

As an example, we briefly demonstrate how to choose $\theta$ in the context of our experiment. Let us take a closer look at the confidence bands (Figure \ref{fig:CIband}) and empirical coverage (Figure \ref{fig:CPthe}).

The fully penalized confidence band with $\theta=1$ (Figure \ref{fig:CI1}) fails to cover the target function near $x=3$ where the function has sharp curvature. When a smaller penalty is used to form the confidence intervals, coverage is improved. The confidence band with $\theta\in\{0,0.05,0.1\}$ (Figures \ref{fig:CI0} and \ref{fig:CIcomp}) covers the truth everywhere across the range of $x$. In fact, all the three choices of $\theta$ yield close-to-nominal coverage level across the range of $x$, as observed in Figure \ref{fig:CPthe}.

Although it has the desired coverage, the unpenalized confidence band with $\theta=0$ (Figure \ref{fig:CI0}) appears extremely wiggly and wide. A slight increase in $\theta$ from 0 to $0.05$ leads to a substantially smoother and narrower confidence band with a well-maintained coverage property as seen in Figure \ref{fig:CIcomp}. When $\theta$ further increases to $0.1$, the resulting confidence band is almost identical to that of $\theta=0.05$, although a modest improvement in smoothness and width can be observed. Figure \ref{fig:CPthe} shows that the resulting coverage suffers near $x=3$, but is nevertheless above $90\%$, and is otherwise as good as that of $\theta=0.05$ and $\theta=0$.

Both $\theta=0.05$ and $\theta=0.1$ generate confidence intervals that perform well at covering the true function without being excessively wide or wiggly. A user would prefer $\theta=0.05$ if a close-to-nominal coverage probability is crucial. On the other hand, $\theta=0.1$ is a reasonable choice if the degraded coverage near $x=3$ is acceptable.

For more general settings, we considered a variety of data generation functions, sampling schemes, and error variances. We repeated the simulation study using two other data generation functions with different degrees of curvature. In addition to equally spaced sample values, we generated sample values from a $5\mathrm{Beta}(0.8,0.8)$ having higher frequency at the lowest and highest values of $x$, and a $5\mathrm{Beta}(1.2,1.2)$ having higher frequency in the middle of the domain. Observing that the model assumption of constant error variance is often violated in real data applications, we let the standard deviation of the error term vary with $x$. Specially, we considered two functions, $\sigma(x)=0.01x+0.075$ and $\sigma(x)=-0.01x+0.125$, where $\sigma(x)$ denotes the standard deviation at the predictor value $x$. The patterns of the simulation results reported in this section repeat themselves in the various contexts mentioned above. The interested reader is referred to the supplementary material \cite[Section 1]{Dai}.

\subsection{Statistical Unfolding Example: Extending to More General Settings}
\label{sub:unfolding}
The proposed method for constructing less biased confidence intervals is not restricted to the penalized spline regression model \eqref{eq:model}. In this section, we apply the proposed method to the statistical unfolding problem described in \cite{KP15}. It is an example of a Poisson inverse problem \citep{AB06,Reiss} which is similar but more complicated than the penalized spline regression for Poisson models. Our point is that the idea of reducing smoothing strength to compute a confidence interval can be applied to general settings where point estimates are inherently biased due to penalization.

We carried out the simulation experiment as described in \citet[Section 5]{KP15}. The expected number of true observations is set to be $10,000$ in our study, while \cite{KP15} also considered two other choices, $1000$ and $20,000$. Following \ppcite{KP15} data analysis, we form the point estimate by using emprical Bayes selection of the regularization parameter $\delta$, which is analogous to the smoothing parameter $\alpha$ in the classical criterion \eqref{eq:cri}.

The proposed method of constructing confidence intervals is implemented as follows. We adopt \ppcite{KP15} approach that uses bootstrap percentile intervals by resampling $200$ i.i.d. observations. For each resampled observation $\boldsymbol y^{*(r)},~r=1,\ldots,200$, we compute a resampled point estimate $\hat{f}^{*(r)}_\theta$ using the reduced regularity strength $\theta\delta^*$, where $\theta$ is a prespecified scalar between 0 and 1, and $\delta^*$ is the regularity strength preselected by empirical Bayes method. The sample $\{\hat{f}^{*(r)}_\theta;~r=1,\ldots,200\}$ is a bootstrap representation of the sampling distribution of $\hat{f}_\theta$ and is then used to form a bootstrap percentile interval for $f$. 

Apart from the MCMC sampler for calculating the posterior mean of $\boldsymbol\beta$ (see Section 2 of the supplement \citep{Dai} for a detailed discussion), we followed exactly the same settings, algorithms and choices of parameters as in \cite{KP15} so that our simulation results can be compared to those reported in \citet[Section 5.2]{KP15}.

\begin{figure}[t!]
    \centering
        \begin{subfigure}[t]{.5\textwidth}
        \centering
        \includegraphics[width=\linewidth]{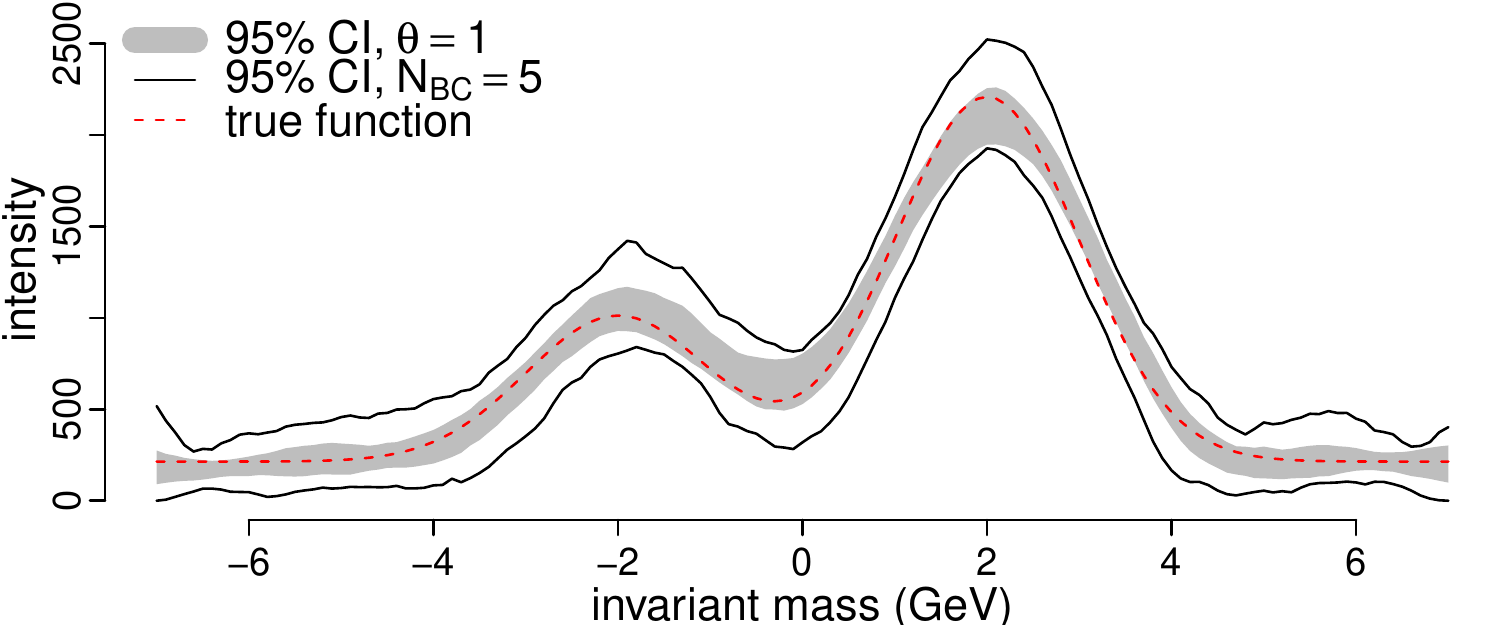}
        \caption{$\theta=1$. Fully penalized.}\label{fig:UF1}
    \end{subfigure}
    ~
    \begin{subfigure}[t]{.5\textwidth}
        \centering
        \includegraphics[width=\linewidth]{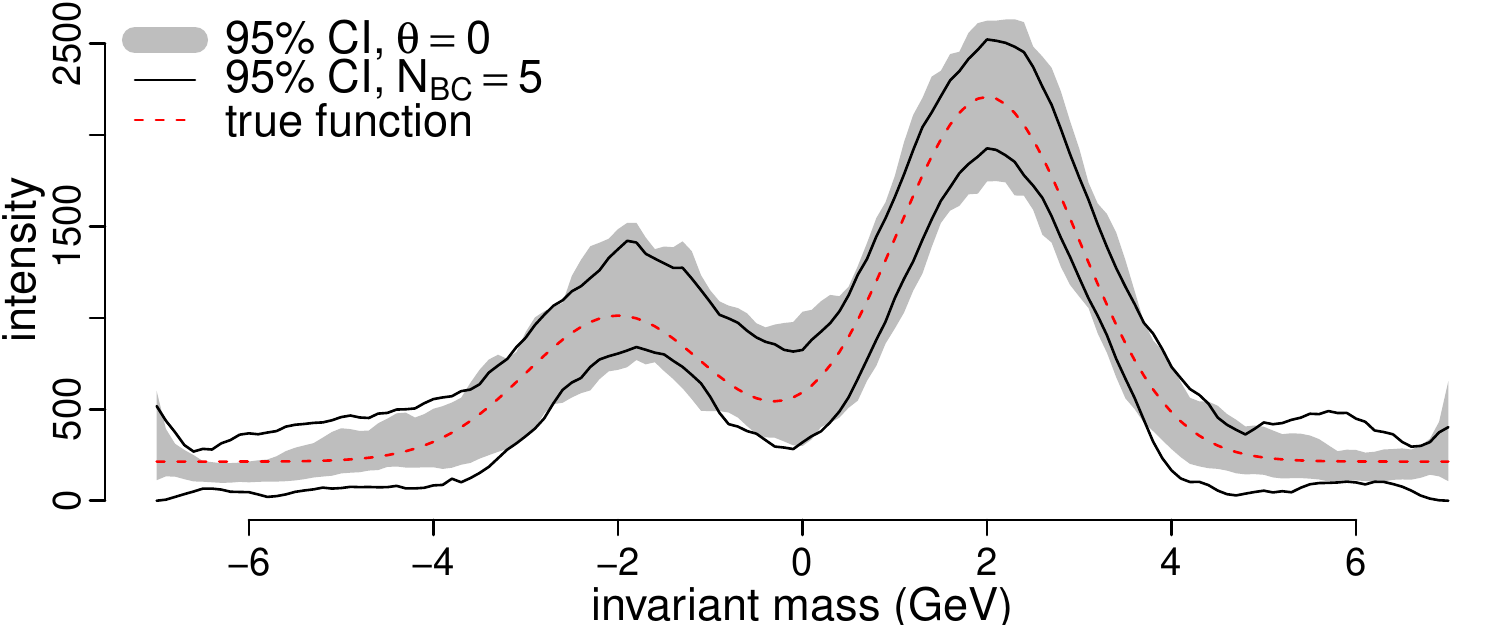}
        \caption{$\theta=0$. Unpenalized.}\label{fig:UF0}
    \end{subfigure}%
    ~ 
    \begin{subfigure}[t]{.5\textwidth}
        \centering
        \includegraphics[width=\linewidth]{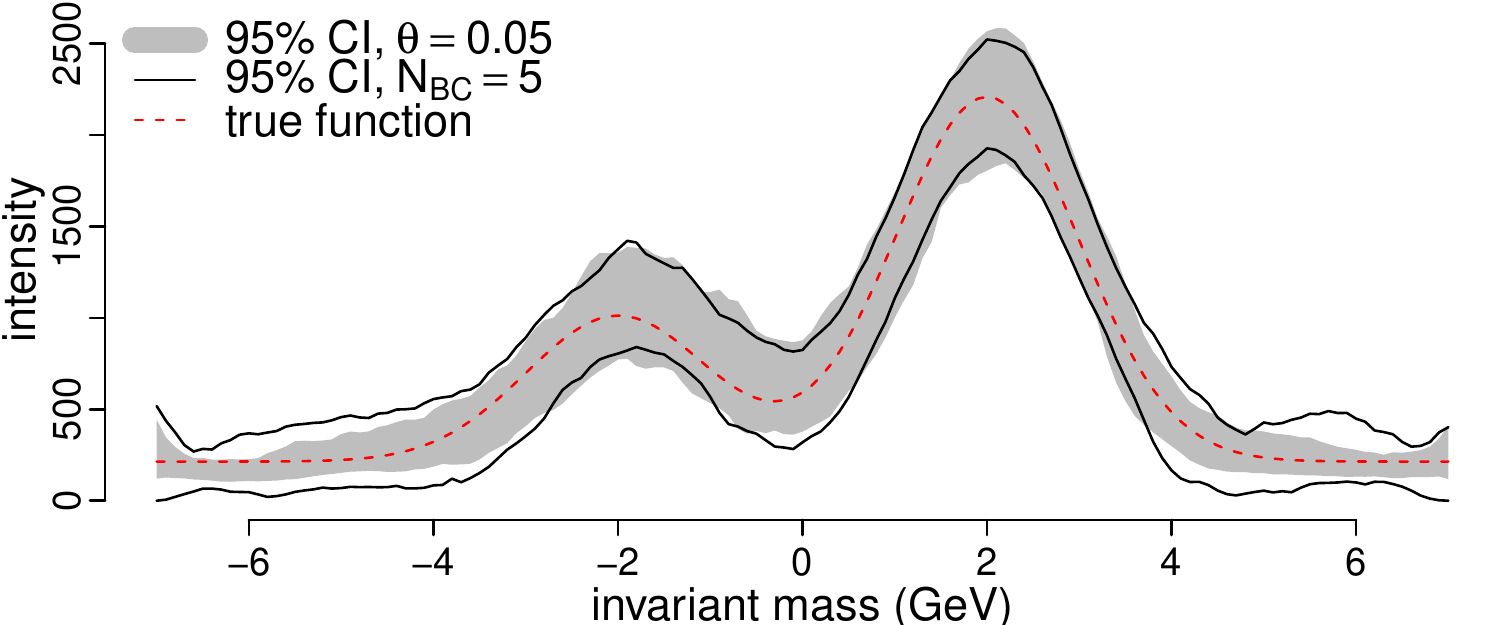}
        \caption{$\theta=0.05$.}\label{fig:UF005}
    \end{subfigure}
    ~
    \begin{subfigure}[t]{.5\textwidth}
        \centering
        \includegraphics[width=\linewidth]{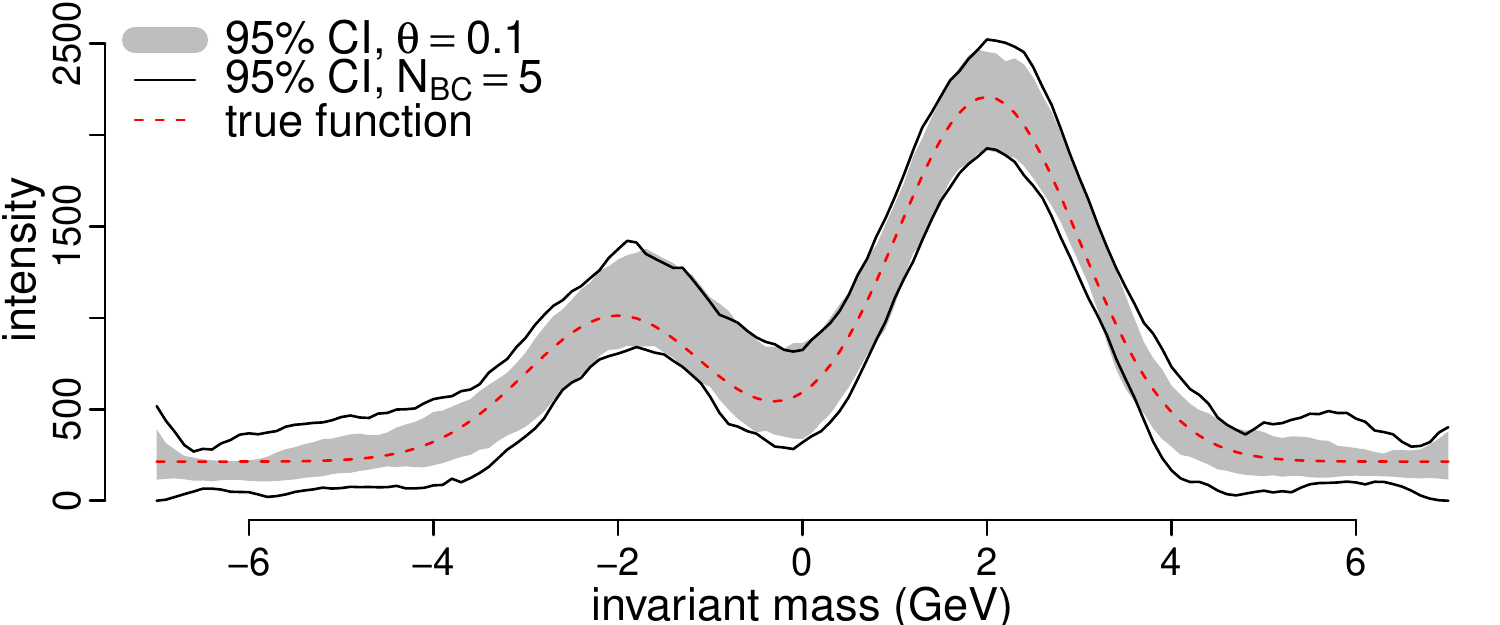}
        \caption{$\theta=0.1$.}\label{fig:UF01}
    \end{subfigure}%
    ~
    \begin{subfigure}[t]{.5\textwidth}
        \centering
        \includegraphics[width=\linewidth]{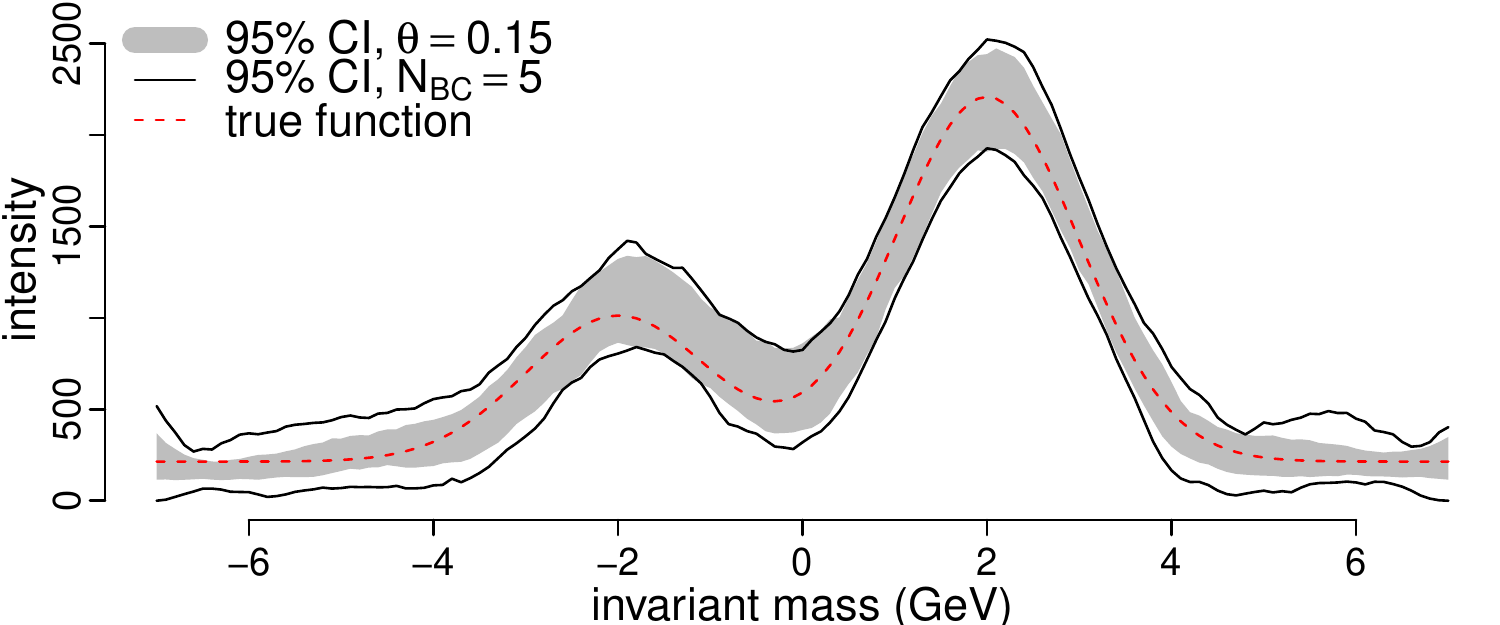}
        \caption{$\theta=0.15$.}\label{fig:UF015}
    \end{subfigure}
    \caption{Statistical unfolding example: Nominal $95\%$ confidence bands produced by the proposed method using different $\theta$ values ($\theta\in\{0,0.05,0.1,0.15\}$) and the fully penalized confidence band ($\theta=1$), with imposed iteratively corrected confidence band using 5 iterations. When $\theta=0.1$ and $0.15$, the resulting confidence band is close to that of the iterative bias-correction method using 5 iterations.}
    \label{fig:UFband}
\end{figure}

Figure \ref{fig:UFband} compares the proposed nominal $95\%$ confidence bands using varying $\theta$ values with \ppcite{KP15} confidence band using 5 iterations. When $\theta$ is as small as $0.15$, the confidence band covers the true function at all points of the predictor values. Substantial improvement is observed over the fully penalized confidence band ($\theta=1$) which is too biased and narrow to cover the true function near $x=2$, where $x$ denotes the predictor.

Note the difference in shape between the proposed confidence band and the iteratively corrected band. The confidence bands produced by the proposed method (Figures \ref{fig:UF0}, \ref{fig:UF005}, \ref{fig:UF01}, \ref{fig:UF015}) are narrow and smooth for $x\in[-7,-4]$ and $x\in[4,7]$ where the true function is flat, and much wider and wigglier near the peaks ($x=-2,2$) and the trough ($x=0$) of the true function. In contrast, \ppcite{KP15} confidence band shows constant width and degree of smoothness across the range of the predictor values. The same pattern is present in Figure 2 of \cite{KP15}. The iteratively corrected interval appears as wide and wiggly as the proposed interval using $\theta=0.1$ (Figure \ref{fig:UF01}) for $x\in[-2.5,2.5]$ where the function shows a moderate to severe degree of curvature, while in flat regions ($x\in[-7,-4]$ and $x\in[4,7]$), the proposed interval using $\theta=0.1$ is noticeably narrower and smoother than the iteratively corrected interval. Even the unpenalized interval, which is usually considered extremely wide and wiggly, has a more desirable shape than the iteratively corrected interval where the function appears flat. In conclusion, in this example the proposed method captures the trend and shape of the true function better than \ppcite{KP15} method.

In constructing bootstrap percentile intervals, the running time to obtain a resampled point estimate, averaged over the 200 bootstrap repetitions, is approximately 1.16 minutes for the proposed method with all the $\theta$ values we considered and 57.85 minutes for \ppcite{KP15} method with 5 iterations, so the proposed method is 50 times faster. \ppcite{KP15} bias-correction approach is computationally expensive because it uses bootstrap iteratively to estimate the bias of the point estimate, which requires executing an MCMC sampler for each bootstrap sample in each bias-correction iteration. Although the user can speed up \ppcite{KP15} algorithm by adopting a faster MCMC sampler, the computational cost will never be comparable to that of the proposed method, which does not involve MCMC.

When applying the proposed method, one simply repeats the point estimation using a smaller regularity strength after the initial analysis. The same software that is used in point estimation can be re-used to get the confidence intervals. \ppcite{KP15} iterative algorithm, by contrast, involves bootstrapping and MCMC sampling. During the process, additional software, time and efforts are demanded. Thus the proposed method is much easier to implement.

\section{Concluding Remarks}
\label{sec:conclude}
We have developed a novel approach to improving the inherently biased confidence intervals for penalized regression splines. The idea is that reducing the smoothing strength leads to less biased spline fits and to confidence intervals with better coverage. When no penalty is applied, the fitted curve is unbiased, and thus the corresponding intervals obtain close-to-nominal coverage.

The unpenalized confidence intervals achieve desirable coverage at the cost of smoothness. To strike a balance between coverage and smoothness, small positive smoothing parameter values should be considered. We observe that a slight increase in the smoothing strength compared to an unpenalized interval gives a significant gain in smoothness while retaining the desired coverage with minimal loss only at predictor values with sharp curvature. With a carefully selected smoothing strength, the proposed confidence intervals perform well at covering the true function without being excessively wide or wiggly.

The proposed method is simple and straightforward to implement. It is easy to select an appropriate amount of reduction in the smoothing strength as the smoothness of the confidence band is fully controlled by the ratio of the smoothing parameter value used to construct the confidence intervals to that of the spline fits. Furthermore, because the proposed method for constructing confidence intervals uses the same machinery as in calculating the fully penalized fits, the same software that is adopted to obtain the fully penalized spline fits can be re-used to get the confidence intervals, so that the computational cost of obtaining a confidence interval is as low as that of point estimation.

The proposed method for constructing less biased confidence intervals is not restricted to the penalized spline regression model with a single predictor. The extension to multivariate situations is straightforward. More importantly, the idea of using a smaller smoothing strength for interval estimation than for point estimation can be applied to penalized likelihood regression with generalized linear models and in other settings where point estimates are inherently biased due to penalty.
\section*{Acknowledgments}
This manuscript originated as a project in a class taught by Jim Hodges, who made helpful suggestions.  The author would also like to thank Galin Jones and Matt Wand for comments and encouragement.

\bibliographystyle{apalike}
\bibliography{projectbib}
\end{document}